\documentclass[10pt,apjl]{emulateapj}
\usepackage{amssymb,graphics} 
\usepackage{multirow} 

\begin{document}

\shorttitle{Identifying Orphan Stream Members from Low-Resolution Spectroscopy}
\shortauthors{Casey et al}

\title{Hunting the Parent of the Orphan Stream: Identifying Stream Members \\ from Low-Resolution Spectroscopy}

\author{Andrew R. Casey\altaffilmark{1,2}, Gary Da Costa\altaffilmark{1}, Stefan C. Keller\altaffilmark{1}, Elizabeth Maunder\altaffilmark{1}}
\altaffiltext{1}{Research School of Astronomy \& Astrophysics, Australian National University, Mount Stromlo Observatory, via Cotter Rd, Weston, ACT 2611, Australia; \email{acasey@mso.anu.edu.au}}
\altaffiltext{2}{Massachusetts Institute of Technology and Kavli Institute for
Astrophysics and Space Research, 77 Massachusetts Avenue,
Cambridge, MA 02139, USA}

\begin{abstract}
We present candidate K-giant members in the Orphan Stream which have been identified from low-resolution data taken with the AAOmega spectrograph on the Anglo-Australian Telescope. From modest S/N spectra and independent cuts in photometry, kinematics, gravity and metallicity we yield self-consistent, highly probable stream members. We find a revised stream distance of $22.5\,\pm\,2.0$\,kpc near the celestial equator, and our kinematic signature peaks at $V_{GSR} = 82.1\,\pm\,1.4$\,km s$^{-1}$. The observed velocity dispersion of our most probable members is consistent with arising from the velocity uncertainties alone. This indicates that at least along this line-of-sight, the Orphan Stream is kinematically cold. Our data indicates an overall stream metallicity of [Fe/H] $= -1.63\,\pm\,0.19$\,dex which is more metal-rich than previously found and unbiased by spectral type. Furthermore, the significant metallicity dispersion displayed by our most probable members, $\sigma(\mbox{[Fe/H]}) = 0.56$\,dex, suggests that the unidentified Orphan Stream parent is a dSph satellite. We highlight likely members for high-resolution spectroscopic follow-up.
\end{abstract}


\section{Introduction}
\label{sec:introduction}

The Milky Way stellar halo has partly formed through the accretion of satellites that are disrupted by tidal forces as they fall into the Galaxy's potential. Stars which were once gravitationally bound to the satellite are distributed along the progenitor's orbit in leading and trailing streams of stars. The velocities of stars in the stream are sensitive to the shape of the dark matter halo, allowing us to constrain the Milky Way potential and reconstruct the formation history of the Galaxy. The level of accreted substructure in the Milky Way has only recently become apparent through multi-band photometric surveys like the Sloan Digital Sky Survey (SDSS). The more prominent of the detectable substructures, like Sagittarius, have been well-studied. One of the more prominent \--- yet less studied \--- substructures is that of the Orphan Stream. 


The Orphan Stream was independently detected by both \citet{Grillmair_2006} and \citet{Belokurov_et-al_2006}, and is distinct from other substructures in the halo. The stream stretches over $60\,^\circ$ in the sky, has a low surface brightness, and a narrow stream width of only $\sim$2\,$^\circ$.  As the name suggests, the parent object largely remains a mystery. The stream extends past the celestial equator \--- outside the SDSS footprint \--- but has not been detected in existing southern surveys \citep{Newberg_et-al_2010}. Whilst the parent system remains elusive, significant effort has been placed on associating the stream with known Milky Way satellites \citep{Zucker_et-al_2006, Fellhaur_et-al_2007,Jin_Lynden_Bell_2007,Sales_et-al_2008}. In contrast, there has been relatively limited observational work on the Orphan Stream itself other than the original discovery papers \citep{Grillmair_2006, Belokurov_et-al_2006, Belokurov_et-al_2007} and the work of \citet{Newberg_et-al_2010}. This is largely to be expected given the absence of deep multi-band photometry in the southern sky and the low total luminosity of the stream. This makes it difficult to reliably separate Orphan Stream members from halo stars. Understanding the full extent of the stream awaits the SkyMapper and Pan-STARRS photometric surveys \citep{Keller_et-al_2007, Hodapp_et-al_2004}.

As \citet{Sales_et-al_2008} point out, there is a natural observational bias towards more massive and recent mergers like Sagittarius. Consequently, the fainter end of this substructure distribution has yet to be fully recovered, or thoroughly examined. Interestingly, there are indications that  some fainter substructures like the Orphan Stream and the Palomar 5 tidal tails \citep{Odenkirchen_et-al_2009} have orbits which seem to be best-fit by Milky Way models with nearly 60\% less mass \citep{Newberg_et-al_2010} than generally reported by \citet{Xue_et-al_2008} and \citet{Koposov_et-al_2010}. Such a discrepancy in the mass of the Milky Way is troublesome. More complete photometric and kinematic maps of these low total luminosity streams may provide the best test as to whether this mass discrepancy is real, or an artefact of incomplete observations. Whilst the full spatial extent of the Orphan Stream remains unknown, we can examine the detailed chemistry of its members, investigate the stream history, and make predictions about the nature of the progenitor.

In this paper we present a detailed, self-consistent analysis to identify K-giant members of the Orphan Stream. Using our selection method we have catalogued the locations of nine highly probable Orphan Stream candidates, all worthy of high-resolution spectroscopic follow up. In the following section we outline our photometric target selection. In \S\ref{sec:observations} we describe the low-resolution spectroscopic observations. The data analysis, including stream identification, is discussed in \S\ref{sec:analysis} and in \S\ref{sec:conclusions} the conclusions, predictions and future work are presented.

\section{Target Selection}
\label{sec:target-selection}

We have targeted K-giant members of the Orphan Stream in order to investigate their detailed chemistry. Because K-giants are difficult to unambiguously detect from photometry alone, low-resolution spectroscopy is required to estimate stellar parameters and determine radial velocities. The Orphan Stream has an extremely low spatial over-density, which makes it difficult to separate stream members from halo stars. However, there is a well described distance gradient along the stream \citep{Belokurov_et-al_2007, Newberg_et-al_2010} which provides an indication on where we should focus our spectroscopic efforts.

The Orphan Stream is closest to us in two locations on the edge of the SDSS footprint: at the celestial equator \citep{Belokurov_et-al_2007}, and along outrigger SEGUE Stripe 1540 \citep{Newberg_et-al_2010}. These two locations are unequivocally the best place to recover bright stream members. We have targeted two fields centered on $(\alpha, \delta) =$ (10:48:15, 00:00:00) and (10:48:15, $-$02:30:00), and employed a combination of colour cuts with the SDSS DR 7 \citep{Abazajian_et-al_2009} data set in order to identify likely K-giants:
\begin{eqnarray}
0.6 <& (g-i)_0 &< 1.7 \\
-15(g-i)_0 + 27 <& g_0 &< -3.75(g-i)_0 + 22.5 \\
15  <& i_0  &< 18 
\end{eqnarray}

Given our colour selection we expect to recover giants and contaminating dwarfs. Although the 2MASS $JHK$ colours can help to separate dwarfs and giants, our target K-giants stars are too faint to be detected in the 2MASS catalogue.

\section{Observations}
\label{sec:observations}

Observations took place on the Anglo-Australian Telescope using the AAOmega spectrograph in April 2009. AAOmega is a fibre-fed, dual beam multi-object spectrograph which is capable of simultaneously observing spectra of 392 (science and sky) targets across a $2\,^\circ$ field of view. We used the 5700\,{\AA} dichroic in combination with the 1000I grating in the red arm, and the 580V grating in the blue arm. This provides a spectral coverage between $800 \leq \lambda \leq 950$\,nm in the red at $\mathcal{R} \approx 4400$, and between $370 \leq \lambda \leq 580$\,nm with a lower spectral resolution of $\mathcal{R} \approx 1300$ in the blue.

The data were reduced using the standard \textsc{2DFDR} reduction pipeline\footnote{http://www.aao.gov.au/2df/aaomega/aaomega\_2dfdr.html}. After flat-fielding, throughput calibration for each fibre was achieved using the intensity of skylines in each fibre. The median flux of dedicated sky fibres was used for sky subtraction, and wavelength calibration was performed using ThAr arc lamp exposures taken between science frames. Three thirty minute science exposures were median-combined to assist with cosmic ray removal. The median S/N obtained in the red arm for our fields is modest at 35 per pixel, although this deteriorates quickly for our fainter targets. With the presence of strong Ca\,\textsc{II} triplet lines in the red arm we are able to ascertain reliable radial velocities and reasonable estimates on overall metallicity \citep[][and references therein]{Starkenburg_et-al_2010}. Our spectral region also includes gravity-sensitive magnesium lines: Mg\,\textsc{I} at 8807\,{\AA}, and the Mg\,\textsc{I}\,b 3$p$-4$s$ triplet lines at $\sim$5178\,{\AA}. As we demonstrate in the next section, these lines are sufficient to discriminate dwarfs from giants even with weak signal.

The blue and red arm spectra were normalised using a third order cubic spline after multiple iterations of outlier clipping. We used defined knot spacings of 20 nm in the red arm, and 5 nm in the blue arm in order to accommodate often poor S/N, and varying strengths of molecular band-heads.

\section{Analysis \& Discussion}
\label{sec:analysis}

We have employed a combination of separate criteria to identify likely Orphan Stream members: kinematics, a giant/dwarf indication from Mg\,\textsc{I} lines, and selecting stars with consistent metallicities derived from both isochrone fitting and the strength of the Ca\,\textsc{II} triplet lines. Each criteria is discussed here separately.

\subsection{Kinematics}
Radial velocities were measured by cross-correlating our normalised spectra against a K-giant synthetic template with a temperature of 4500\,K, $\log{g}$ = 1.5 and $[\mbox{M/H}] = -1.5$ across the range $845 \leq \lambda \leq 870$\,nm. Heliocentric velocities were translated to the galactic rest frame by adopting the local standard of rest velocity as 220\,km s$^{-1}$ towards $(l, b) = (53\,^\circ, 25\,^\circ)$ \citep{Kerr_Lynden-Bell_1986, Mihalas_Binney_1981}\footnote{Where $V_{GSR} = V_{HELIO} + 220\sin{l}\cos{b} + 16.5\times[\sin{b}\sin{25^\circ} + \cos{b}\cos{25^\circ}\cos{(l - 53^\circ)}]$}.

\begin{figure}[h]
	\includegraphics[width=\columnwidth]{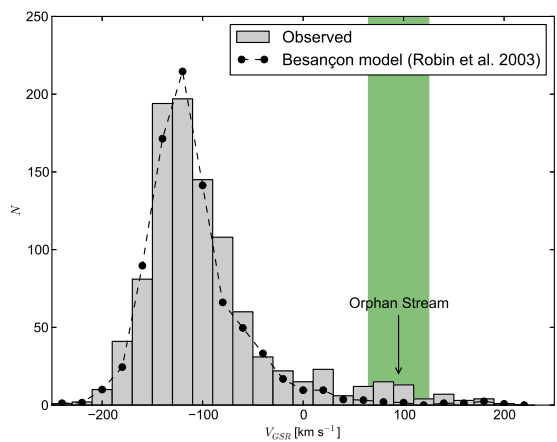}
	\caption{Galactocentric rest frame velocities for stars in both our observed fields (grey), and predicted Besan\c{c}on velocities which have been scaled to match our observed sample size. The expected kinematic signature from \citet{Newberg_et-al_2010} for the Orphan Stream is highlighted, as is our kinematic selection window (green).}
	\label{fig:velocities}
\end{figure}

Figure \ref{fig:velocities} shows a histogram of our galactocentric velocities, compared to the predicted smooth line-of-sight velocity distribution for this region from the Besan\c{c}on model \citep{Robin_et-al_2003}. We have selected particles from the Besan\c{c}on model using the same criteria outlined in \S\ref{sec:target-selection} after employing the \citet{Jordi_et-al_2006} colour transformations. It is clear that our target selection has yielded mostly nearby disk dwarf stars with $V_{GSR} \approx -120$\,km s$^{-1}$.

In a nearby ($\Delta\Lambda_{Orphan} \sim 4\,^\circ$) region of the stream, \citet{Newberg_et-al_2010} detected the Orphan Stream with a $V_{GSR} = 101.4$ km s$^{-1}$ from BHB stars. Differences in accounting for the local standard of rest between this work and \citet{Newberg_et-al_2010} means that this corresponds to approximately 95\,km s$^{-1}$ on our $V_{GSR}$ scale.  This is discussed further in \S\ref{sec:newberg}. The expected Orphan Stream kinematic peak is labelled in Figure \ref{fig:velocities}. There is no obvious sharp kinematic peak representative of the Orphan Stream in our sample. From kinematics alone, our targets appears largely indistinguishable from a smooth halo distribution. To isolate potential Orphan Stream members we have nominated a relatively wide selection criteria between $65 \leq V_{GSR} \leq 125$ km s$^{-1}$ (shown in Figure \ref{fig:velocities}), which yields 28 Orphan Stream candidates. The typical uncertainty in our velocities is $\pm{}5.0$\,km s$^{-1}$.

\subsection{Dwarf/Giant Discrimination}
\label{sec:dwarf-giant}

We have measured the equivalent width of the gravity-sensitive Mg\,\textsc{I} line at 8807 \AA{} to distinguish dwarfs from giants \citep{Battaglia_Starkenburg_2012}. At a given temperature (or $g - r$) and metallicity, giant stars present narrower Mg\,\textsc{I} absorption lines than their dwarf counterparts. Given the target selection, our sample is likely to contain many more dwarfs than giants (e.g. see \citet{Casey_et-al_2012} where a similar colour selection was employed). In some cases no Mg\,\textsc{I} 8807 \AA{} line was apparent, so an upper limit was estimated based on the S/N of the spectra. In these cases the candidate was considered a ``non-dwarf" because we cannot exclusively rule out a metal-poor sub-giant with this criteria alone. For these purposes we are only looking for a simple indication as to whether a star is likely a dwarf or not. 

\begin{figure}[h]
	\includegraphics[width=\columnwidth]{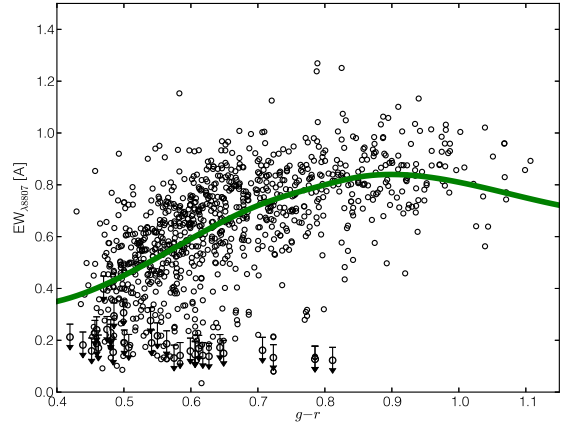}
	\caption{SDSS $g - r$ against the measured equivalent width of the Mg\,\textsc{I} transition at 8807\,\AA{}. Dwarf contaminants occupy the more populous upper branch. Our separation line between dwarfs and giants is shown in green.}
	\label{fig:ew-mg}
\end{figure}

Figure \ref{fig:ew-mg} illustrates the trend with $EW_{\lambda8807}$ against SDSS de-reddened\footnote{All magnitudes presented in this letter are de-reddened using the \citet{Schlegel_Finkbeiner_Davis_1998} dust maps.} $g - r$, illustrating the dominant upper dwarf branch we wish to exclude. Giant stars populate the lower, sparser branch. A separation line has been adopted to distinguish dwarfs from giants, and is shown in Figure \ref{fig:ew-mg}. If we were to place this line higher, the total number of true giant stars may increase, but the dwarf contamination rate will rise dramatically. A compromise must be made between the rate of giant recoverability and the dwarf contamination. Our  dwarf/giant separation line lies just below the main dwarf population. On its own, this dwarf/giant separation line would typically result in far too many dwarf contaminants. However, we are  employing selections on multiple observables (kinematics, metallicity, proper motions) in order to refine our Orphan Stream giant sample. 

This dwarf/giant separation method was also employed using the total equivalent width of the Mg\,\textsc{I}\,b triplet lines. Both analyses were entirely consistent with each other: essentially the same candidate list was found using both techniques. However, given slightly poorer signal at the Mg\,\textsc{I}\,b triplet, we were forced to adopt many more upper limits than when using the 8807\,{\AA} line. Because we classify all upper limits as being ``non-dwarfs" (i.e. potential giants), we deduced a slightly larger candidate sample for the Mg\,\textsc{I}\,b analysis, which was primarily populated by upper limits. In conclusion, we found the Mg\,\textsc{I} line at 8807\,{\AA} appeared to be a more consistent dwarf discriminant given our weak S/N \--- particularly for our fainter stars. Thus, we have used the 8807\,{\AA} Mg\,\textsc{I} selection throughout the rest of our analysis.

Our dwarf/giant separation line in Figure \ref{fig:ew-mg} yields 425 potential giants. Upon taking the intersection of our kinematic and gravity selections, we find 20 stars that appear to be likely Orphan Stream giants.

\subsection{Metallicities}
\label{sec:metallicities}

We have measured the metallicities for the stars that meet our kinematic and surface gravity criteria in two ways: with the strength of their Ca\,\textsc{II} triplet lines, and by isochrone-fitting. After correcting for luminosity, the equivalent width of the Ca\,\textsc{II} triplet lines provide a good indication of the overall metallicity of a RGB star \citep{Amandroff_Da_Costa_1991}. We have employed the \citet{Starkenburg_et-al_2010} relationship and corrected for luminosity in $g$ against the horizontal branch magnitude at $g_{HB}$ = 17.1 \citep{Newberg_et-al_2010}. Strictly speaking, the Ca\,\textsc{II}\---[Fe/H] calibration is only valid for stars brighter than the horizontal branch, although the relationship only becomes significantly inappropriate near $g \-- g_{HB} \sim +1$ \citep{Saviane_et-al_2012}. Many of our candidates are fainter than this valid luminosity range, and therefore they should not be excluded solely because of their derived metallicities, as these could be uncertain. Stars fainter than $g_{HB}$ will have slightly lower metallicities than predicted by our Ca\,\textsc{II}\---[Fe/H] relationship, and for these stars we will only use metallicities to assign a relative qualitative likelihood for stream membership.

\begin{figure}[t!]
	\includegraphics[width=\columnwidth]{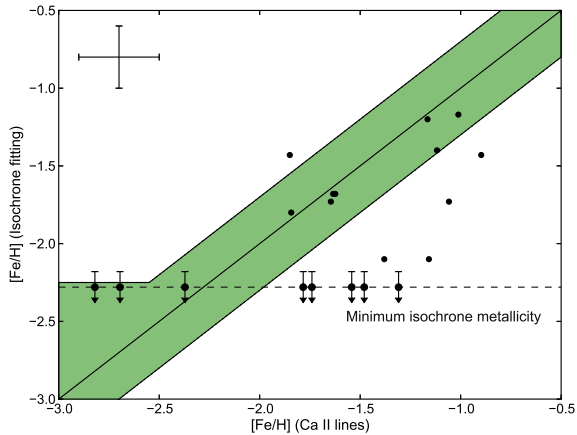}
	\caption{Metallicities from the Ca\,\textsc{II} triplet lines versus those found from fitting isochrones to the 20 stars that meet our kinematic and surface gravity criteria. Both abundance determinations imply these stars are RGB members of the Orphan Stream at a distance of $\sim$21.4\,kpc \citep{Newberg_et-al_2010}. Consistency between these methods indicates highly likely stream membership (shaded region). The minimum isochrone [Fe/H], and a representative uncertainty of 0.2\,dex for abundance measurements is shown.}
	\label{fig:feh}
\end{figure}

Given a distance estimate to the Orphan Stream, we can also deduce a star's metallicity through isochrone fitting. We have used a 10\,Gyr \citet{Girardi_et-al_2008} isochrone at 21.4\,kpc \citep{Newberg_et-al_2010} and found metallicities for all 20 likely stream members from their best-fitting isochrone. Derived metallicities from Ca\,\textsc{II} line strengths and isochrone fitting that are consistent (within $\pm0.3$\,dex) indicates these measurements are reliable, and that these stars are indeed at a distance of $\sim$21.4\,kpc. We find ten highly likely stream members with consistently derived metallicities. They fall within the shaded region illustrated in Figure \ref{fig:feh}. 

\begin{figure}[t!]
	\includegraphics[width=\columnwidth]{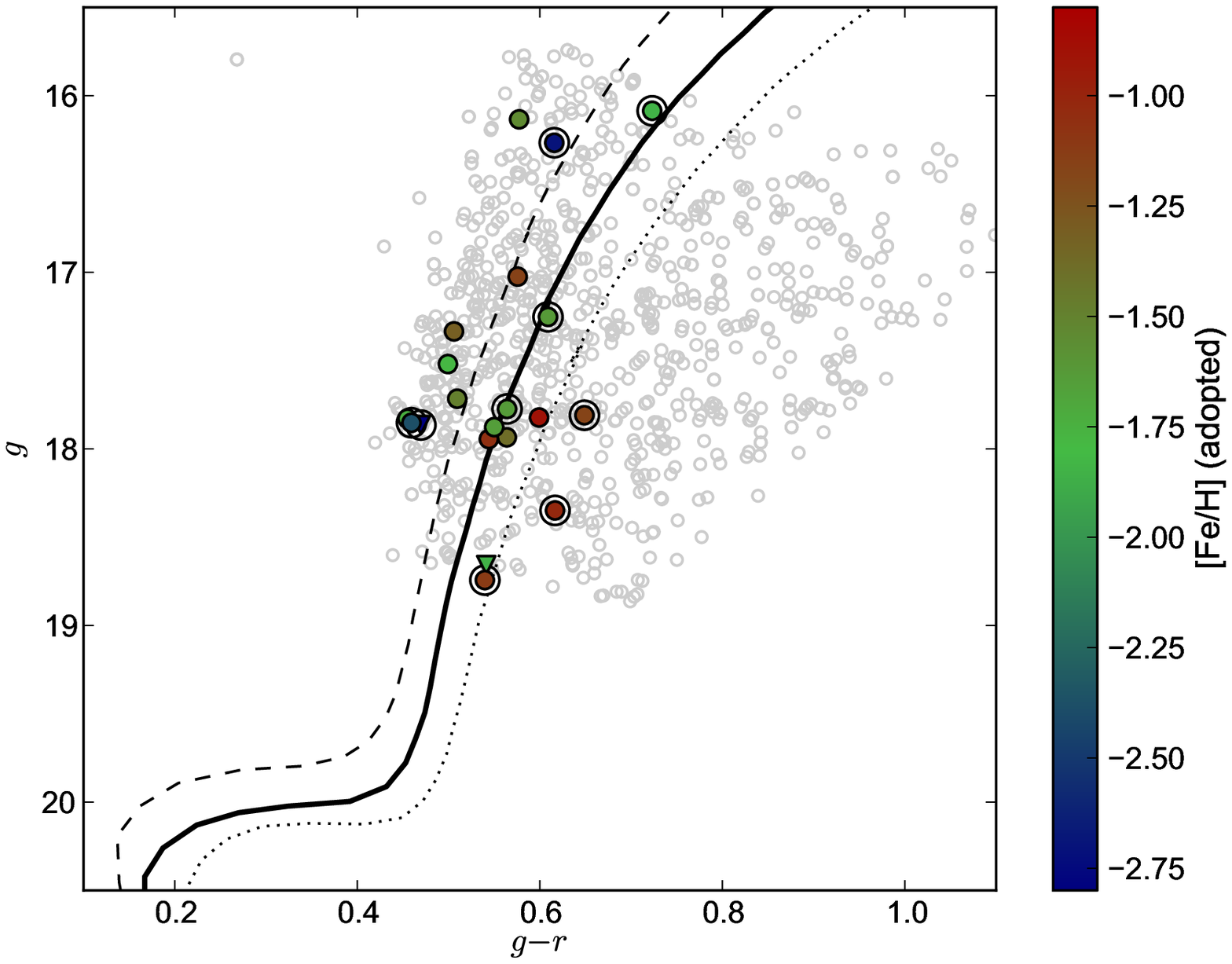}
	\caption{Color magnitude diagram showing our observed candidates (grey). Observations fulfilling kinematic and gravity cuts are colored by their metallicity, and those with upper limits for surface gravity are marked as triangles ($\triangledown$). Highly probable stream members (see text) are circled. Relevant 10\,Gyr \citet{Girardi_et-al_2008} isochrones at $[\mbox{Fe/H}] = -1.5$ (dotted), $-2.0$ (dashed) at 21.4\,kpc \citep{Newberg_et-al_2010} are shown, as well as our best-fitting 10\,Gyr isochrone of $[\mbox{Fe/H}] = -1.63$ at 22.5\,kpc (solid).}
	\label{fig:cmd}
\end{figure}

A final metallicity value for each star has been adopted based on the quality of our [Fe/H] measurements. These values are tabulated in Table \ref{tab:oss-members}. From our highly likely stream members we find an overall stream metallicity of $[\mbox{Fe/H}] = -1.63$ with a dispersion of $\sigma = 0.56$ dex. This abundance spread is larger than typically seen in globular cluster stars and is more representative of the chemical spread seen in dSph satellites \citep[e.g.,][]{Frebel_Norris_2011}.

\setlength{\tabcolsep}{2.4pt}

\begin{table*}[t!]\footnotesize
\caption{Identified Orphan Stream Candidates\label{tab:oss-members}}
\begin{tabular*}{\textwidth}{lccccrcrcrrccccc}
\hline
\hline
Star & $\alpha$ & $\delta$ & $g$ & $g - r$ & \multicolumn{2}{c}{$\mu_\alpha$} & \multicolumn{2}{c}{$\mu_\delta$} & \multicolumn{2}{c}{$V_{GSR}$} & $EW_{\lambda8807}$ & [Fe/H]$_{Ca}$ & [Fe/H]$_{iso}$ & [Fe/H]\footnote{Final adopted [Fe/H] value based on quality of two metallicity measurements.} & Stream \\
Name & (J2000) & (J2000) & & & \multicolumn{2}{c}{(mas yr$^{-1}$)} & \multicolumn{2}{c}{(mas yr$^{-1}$)} & \multicolumn{2}{c}{(km s$^{-1}$)} & (m\AA{}) & (dex) & (dex) & (dex) & Prob. \\
\hline
OSS--1  & 10:46:21.9 & $+$00:43:21.8 & 17.52 & 0.50 & 4.1&$\pm$ 4.5 & $-$34.0 &$\pm$ 4.5 & 73.3 &$\pm$ \phn9.3 & 0.273 & --1.78 &$<$--2.28\phn\,& --1.78 & Low \\
OSS--2  & 10:46:29.3 & $-$00:19:38.5 & 17.77 & 0.56 & $-$1.7 &$\pm$ 4.3 & $-$2.2 &$\pm$ 4.3 & 78.4 &$\pm$ \phn5.2 & 0.126 & --1.63 & --1.68  & --1.63 & High \\
OSS--3  & 10:46:50.4 & $-$00:13:15.6 & 17.33 & 0.51 & 1.8 &$\pm$ 4.3 & $-$4.6 &$\pm$ 4.3 & 77.0 &$\pm$ \phn4.0 & 0.416 & --1.31 &$<$--2.28\phn\,& --1.31 & Low \\
OSS--4  & 10:47:06.1 & $-$01:56:03.9 & 18.74 & 0.54 & $-$6.3 &$\pm$ 4.9 & 4.6 &$\pm$ 4.9 & 74.9 &$\pm$    17.6 & 0.452 &:--1.12\footnote{Sufficiently fainter than $g_{HB}$ to qualify this measurement as uncertain.}& --1.40  & --1.40 & High \\
OSS--5  & 10:47:15.0 & $-$03:15:03.9 & 18.66 & 0.54 & $-$8.2 &$\pm$ 5.2 & 1.7 &$\pm$ 5.2 & 109.5 &$\pm$ \phn9.0 &$<$0.19&:--1.85\textsuperscript{b}& --1.43  & --1.43 & Medium \\
OSS--6  & 10:47:17.6 & $+$00:25:07.7 & 16.09 & 0.72 & $-$0.8 &$\pm$ 4.0 & $-$5.2 &$\pm$ 4.2 & 79.2 &$\pm$ \phn3.3 & 0.212 & --1.84 & --1.80  & --1.84 & High \\
OSS--7  & 10:47:29.1 & $-$02:02:22.6 & 17.86 & 0.47 & \multicolumn{2}{c}{\nodata} & \multicolumn{2}{c}{\nodata}  & 93.2 &$\pm$    29.8 &$<$0.40& --2.82 &$<$--2.28\phn\,& --2.82 & High \\
OSS--8  & 10:47:30.1 & $-$00:01:24.5 & 17.25 & 0.61 & $-$4.0 &$\pm$ 4.2 & $-$5.2 &$\pm$ 4.2 & 83.6 &$\pm$ \phn3.5 & 0.123 & --1.62 & --1.68  & --1.62 & High \\
OSS--9  & 10:48:20.9 & $+$00:26:34.4 & 17.88 & 0.55 & $-$8.1 &$\pm$ 4.3 & $-$5.4 &$\pm$ 4.3 & 118.9 &$\pm$    11.7 & 0.467 & --1.65 & --1.73  & --1.65 & High \\
OSS--10 & 10:48:27.8 & $+$00:55:24.0 & 17.72 & 0.51 & $-$14.4 &$\pm$ 4.6 & $-$5.3 &$\pm$ 4.6 & 124.5 &$\pm$ \phn6.7 & 0.182 & --1.48 &$<$--2.28\phn\,& --1.48 & Low \\
OSS--11 & 10:48:31.9 & $+$00:03:35.7 & 17.02 & 0.58 & $-$3.6 &$\pm$ 4.1 & $-$7.7 &$\pm$ 4.1  & 105.1 &$\pm$ \phn5.1 & 0.234 & --1.12 & --2.10  & --1.12 & Low \\
OSS--12 & 10:48:44.4 & $-$02:53:08.8 & 18.35 & 0.62 & $-$3.4 &$\pm$ 4.7 & $-$1.8 &$\pm$ 4.7 & 108.2 &$\pm$ \phn9.0 & 0.183 &:--1.01\textsuperscript{b}& --1.17  & --1.17 & High \\
OSS--13 & 10:48:46.9 & $-$00:32:27.8 & 17.85 & 0.46 & $-$28.4 &$\pm$ 4.8 & $-$12.3 &$\pm$ 4.8 & 109.3 &$\pm$ \phn8.1 & 0.324 & --2.37 &$<$--2.28\phn\,& --2.37 & Medium \\
OSS--14 & 10:49:08.3 & $+$00:02:00.2 & 16.27 & 0.62 & 4.9 &$\pm$ 4.0 & $-$6.0 &$\pm$ 4.0& 81.5 &$\pm$ \phn4.6 & 0.034 & --2.70 &$<$--2.28\phn\,& --2.70 & High \\
OSS--15 & 10:49:13.4 & $+$00:04:03.8 & 17.83 & 0.46 & 3.4 &$\pm$ 4.7 & $-$6.8 &$\pm$ 4.7 & 65.3 &$\pm$ \phn5.4 & 0.252 & --1.74 & $<$--2.28\phn\,& --1.74 & Medium \\
OSS--16 & 10:50:13.1 & $+$00:33:52.7 & 16.13 & 0.58 & $-$3.4 &$\pm$ 4.0 & $-$7.7 &$\pm$ 4.0 & 94.7 &$\pm$ \phn5.1 & 0.391 & --1.54 &$<$--2.28\phn\,& --1.54 & Low \\
OSS--17 & 10:50:24.2 & $-$01:49:05.4 & 17.94 & 0.54 & 3.4 &$\pm$ 4.6 & $-$2.2 &$\pm$ 4.6 & 109.9 &$\pm$    25.4 & 0.151 & --1.06 & --1.73  & --1.06 & Low \\
OSS--18 & 10:50:33.8 & $+$00:12:19.1 & 17.82 & 0.60 & $-$7.4 &$\pm$ 5.0 & $-$3.3 &$\pm$ 5.0 & 97.5 &$\pm$ \phn5.9 & 0.596 & --0.90 & --1.43  & --0.90 & Medium \\
OSS--19 & 10:51:19.7 & $+$00:05:15.5 & 17.81 & 0.65 & 4.2 &$\pm$ 4.7 & $-$11.8 &$\pm$ 4.7 & 82.7 &$\pm$ \phn5.0 & 0.198 & --1.16 & --1.20  & --1.16 & High \\
OSS--20 & 10:51:35.4 & $+$00:00:46.4 & 17.93 & 0.56 & --0.5 &$\pm$ 4.5 & $-$3.0 &$\pm$ 4.5 & 66.7 &$\pm$ \phn8.7 & 0.128 & :--1.38 & --2.10 & --2.10 & Medium \\

\hline
\hline
\end{tabular*}
\end{table*}

\break
\subsection{Proper Motions \& Distances}

We have found proper motions for 19 of our candidates in the PPMXL proper motion catalogue \citep{Roeser_et-al_2010}. One highly probably candidate (OSS--13 in Table \ref{tab:oss-members}) has a listed proper motion that is different from that of the other nine highly likely members at the $6\,\sigma$ level. Consequently, we have reduced the membership likelihood of this star from ``High" to ``Medium". Given the uncertainties in proper motions, we cannot reliably alter the membership probability for any other candidate.

Since our Orphan Stream giants cover a wide evolutionary range along the giant branch (Figure \ref{fig:cmd}), we are in a good position to revise the distance estimate to the stream. Given a 10\,Gyr \citet{Girardi_et-al_2008} isochrone at $[\mbox{Fe/H}] = -1.63$, we find a best-fitting distance to the stream of $22.5 \pm 2.0$\,kpc at $(l, b) = (250\,^\circ,\,50\,^\circ)$. This isochrone is shown in Figure \ref{fig:cmd}. Our derived distance is in reasonably good agreement with the measurement of $21.4 \pm 1.0$\,kpc independently deduced by \citet{Grillmair_2006} and \citet{Newberg_et-al_2010}.

\subsection{Comparison with \citet{Newberg_et-al_2010}}
\label{sec:newberg}
\begin{figure}[h!]
	\includegraphics[width=\columnwidth]{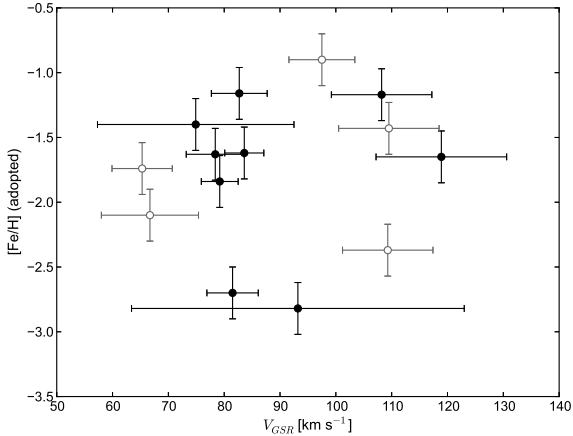}
	\caption{Galactocentric velocities and adopted metallicities for the highest likely Orphan Stream members (black $\bullet$), and those with probabilities assigned as ``Medium'' (grey $\circ$; see text).}
	\label{fig:vgsr-feh}
\end{figure}

\citet{Newberg_et-al_2010} traced the Orphan Stream using BHB stars selected from the SEGUE survey, allowing them to derive an orbit for the stream and make a strong prediction for the location of the undiscovered progenitor. Their closest stream detection to this study is at $\Lambda_{Orphan} = 18.4\,^\circ$, approximately $\Delta\Lambda_{Orphan} \sim 4\,^\circ$ away from our fields. At this location, \citet{Newberg_et-al_2010} found the velocity of the stream to be $V_{GSR} = 101.4 \pm 8.9$\,km s$^{-1}$ based on 12 BHB stars. We note that this is $\sim95$\,km s$^{-1}$ on our scale, given the differences in accounting for the local standard of rest. The velocities and metallicities of our `High' and `Medium' probability candidates are illustrated in Figure \ref{fig:vgsr-feh}. Although we recover some candidates with velocities up to $V_{GSR} \sim 110$\,km s$^{-1}$, our kinematic distribution peaks near $V_{GSR} \sim 85$\,km s$^{-1}$, roughly 10\,km s$^{-1}$ lower than that of \citet{Newberg_et-al_2010}.

There is a known velocity gradient along the Orphan Stream which can account for this discrepancy. As $\Lambda_{Orphan}$ increases towards the edge of the SDSS boundary, galactocentric velocity quickly decreases. For the Orphan Stream detection in the outrigger SEGUE Strip 1540 at $\Lambda_{Orphan} = 36\,^\circ$, \citet{Newberg_et-al_2010} find $V_{GSR} = 38$\,km s$^{-1}$. This work presents likely Orphan Stream K-giant candidates at $\Lambda_{Orphan} \sim 23\,^\circ$. Given the velocity gradient reported by \citet{Newberg_et-al_2010}, a galactocentric velocity of $80-85$\,km s$^{-1}$ (on our scale) is perfectly reasonable. We note that since the velocities of BHB stars can have significant uncertainties, it was practical for us to assume a wide initial selection in kinematics to identify potential members. 

\begin{figure}[h!]
	\includegraphics[width=\columnwidth]{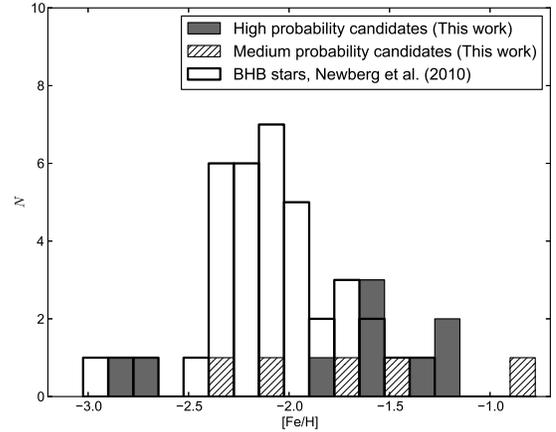}
	\caption{Observed metallicity distribution function for the Orphan Stream candidates identified here, with comparisons to the distribution found by \citet{Newberg_et-al_2010} from BHB stars.}
	\label{fig:newberg-feh}
\end{figure}

The adopted metallicities of our Orphan Stream candidates are generally higher than those found by \citet{Newberg_et-al_2010}. Our highly likely stream members have a mean metallicity of [Fe/H] = $-1.63$, with a dispersion of $\sigma = 0.56$\,dex. As illustrated in Figures \ref{fig:vgsr-feh} and \ref{fig:newberg-feh}, there are two very metal-poor candidates which largely drive this dispersion, but we have no reason to suspect they are non-members. The \citet{Newberg_et-al_2010} sample contains 37 BHB stars identified over a 60\,$^\circ$ arc on the sky, and has a peak metallicity at [Fe/H] = $-2.10 \pm 0.10$. The closest detection bin in the \citet{Newberg_et-al_2010} sample was the most populous, yielding 7 BHB stars. For comparison, we identify 9 giant stars across $\sim{}4\,^\circ$. Given BHB stars are known to trace a somewhat more metal-poor population, and we are calculating statistics with marginal sample sizes, we conclude that the accuracy of these two metallicity distributions are not mutually exclusive. It is entirely possible that we are sampling the same distribution, but a larger sample size is required.

\section{Conclusions}
\label{sec:conclusions}

We have presented a detailed analysis to isolate individual Orphan Stream K-giants from low-resolution spectroscopy using a combination of photometric, kinematic, gravity, metallicity, and proper motion information. Although each individual criterion is likely to induce some level of contamination, their intersection reveals nine highly probable, self-consistent, Orphan Stream K-giants.  We deduce a median stream metallicity of $[\mbox{Fe/H}] = -1.63 \pm 0.19$ and find an intrinsically wide metallicity spread of $\sigma = 0.56$\,dex, indicative of a dSph origin. Unlike other stellar tracers, K-type giants can exist at all metallicities, hence our derived metallicity spread is likely representative of the true stream metallicity distribution function. Recall that the metallicity determination was performed after kinematic and gravity cuts, and three of our most probable members lay perfectly on a 10\,Gyr isochrone of $[\mbox{Fe/H}] = -1.63$. However, it is clear that more data is required to fully characterize the stream metallicity distribution function. Our data indicate a distance to the stream of $22.5 \pm 2.0$\,kpc at $(l, b) = (250\,^\circ,\,50\,^\circ)$, in agreement with that deduced by \citet{Grillmair_2006} and \citet{Newberg_et-al_2010}.

Given the stream orbit derived by \citet{Newberg_et-al_2010}, they excluded all possible known halo objects except for the dissolved star cluster, Segue 1. \citet{Simon_et-al_2011} obtained spectroscopy for six members in Segue 1 and found an extremely wide metallicity dispersion: from $<-3.4$ to $-1.63$ dex. On the basis of the extremely low metallicity in the cluster and the wide chemical dispersion, they conclude that Segue 1 is a disrupted dwarf spheroidal galaxy. Although the data presented here indicates the Orphan Stream progenitor is a disrupted dwarf spheroidal galaxy, we cannot reliably associate Segue 1 as the parent without additional observational data. 

If the Orphan Stream continues through SEGUE Stripe 1540 at $(l, b) = (271\,^\circ,\,38\,^\circ)$ as \citet{Newberg_et-al_2010} found, then the stream is even closer there than in the region analysed here. Thus, if our observations and analyses are repeated at $(271\,^\circ,\,38\,^\circ)$, we predict K-giant stream members of brighter apparent magnitude will be recovered. 

Using a maximum-likelihood estimation we find the stream velocity at $(l, b) = (250\,^\circ, 50\,^\circ)$ from nine stars to be $V_{GSR} = 85.3 \pm 4.4$\,km s$^{-1}$ and the dispersion to be $6.5 \pm 7.0$\,km s$^{-1}$. If we exclude three stars with low signal-to-noise \--- and hence large ($> 10$\,km s$^{-1}$) velocity uncertainties \--- the peak occurs at $82.1 \pm 1.4$\,km s$^{-1}$ and the intrinsic dispersion is found to be $0.2\,\pm\,3.1$\,km s$^{-1}$. Hence, the observed stream dispersion is dominated by the velocity uncertainties, indicating that the intrinsic dispersion is small. 

The K-giants presented here can provide great insight
\begin{minipage}{\columnwidth}
into the chemistry and history of the Orphan Stream. High-resolution spectroscopic observations have been taken for some of our highly probable members and a detailed chemical analysis will be presented in a forthcoming paper (Casey et al., in preparation). Detailed chemical abundances can help determine both the nature of the progenitor before it is discovered, and allows us to compare peculiar chemical signatures with those of the known Milky Way satellites in order to associate likely parents. However, at least for the moment, the Orphan Stream remains appropriately named.
\end{minipage}

\acknowledgements
The authors would like to thank the anonymous referee for their constructive comments which improved this paper. ARC acknowledges the financial support through the Australian Research Council Laureate Fellowship 0992131, and from the Australian Prime Minister's Endeavour Award Research Fellowship, which has facilitated his research at MIT. SK and GDaC acknowledge the financial support from the Australian Research Council through Discovery Programs DP0878137 and DP120101237. GDaC is also grateful for the support received during an extended visit to the Institute of Astronomy, University of Cambridge, during which this work was completed.

Funding for the SDSS and SDSS-II has been provided by the Alfred P. Sloan Foundation, the Participating Institutions, the National Science Foundation, the U.S. Department of Energy, the National Aeronautics and Space Administration, the Japanese Monbukagakusho, the Max Planck Society, and the Higher Education Funding Council for England. The SDSS Web Site is http://www.sdss.org/.

The SDSS is managed by the Astrophysical Research Consortium for the Participating Institutions. The Participating Institutions are the American Museum of Natural History, Astrophysical Institute Potsdam, University of Basel, University of Cambridge, Case Western Reserve University, University of Chicago, Drexel University, Fermilab, the Institute for Advanced Study, the Japan Participation Group, Johns Hopkins University, the Joint Institute for Nuclear Astrophysics, the Kavli Institute for Particle Astrophysics and Cosmology, the Korean Scientist Group, the Chinese Academy of Sciences (LAMOST), Los Alamos National Laboratory, the Max-Planck-Institute for Astronomy (MPIA), the Max-Planck-Institute for Astrophysics (MPA), New Mexico State University, Ohio State University, University of Pittsburgh, University of Portsmouth, Princeton University, the United States Naval Observatory, and the University of Washington.


\vspace{2em}

\begin{thebibliography}{29}
\expandafter\ifx\csname natexlab\endcsname\relax\def\natexlab#1{#1}\fi

\bibitem[{{Abazajian} {et~al.}(2009){Abazajian}, {Adelman-McCarthy},
  {Ag{\"u}eros}, {Allam}, {Allende Prieto}, {An}, {Anderson}, {Anderson},
  {Annis}, {Bahcall}, \& et~al.}]{Abazajian_et-al_2009}
{Abazajian}, K.~N., {et~al.} 2009, \apjs, 182, 543

\bibitem[{{Armandroff} \& {Da Costa}(1991)}]{Amandroff_Da_Costa_1991}
{Armandroff}, T.~E., \& {Da Costa}, G.~S. 1991, \aj, 101, 1329

\bibitem[{{Battaglia} \& {Starkenburg}(2012)}]{Battaglia_Starkenburg_2012}
{Battaglia}, G., \& {Starkenburg}, E. 2012, \aap, 539, A123

\bibitem[{{Belokurov} {et~al.}(2006){Belokurov}, {Zucker}, {Evans}, {Gilmore},
  {Vidrih}, {Bramich}, {Newberg}, {Wyse}, {Irwin}, {Fellhauer}, {Hewett},
  {Walton}, {Wilkinson}, {Cole}, {Yanny}, {Rockosi}, {Beers}, {Bell},
  {Brinkmann}, {Ivezi{\'c}}, \& {Lupton}}]{Belokurov_et-al_2006}
{Belokurov}, V., {et~al.} 2006, \apjl, 642, L137

\bibitem[{{Belokurov} {et~al.}(2007){Belokurov}, {Evans}, {Irwin},
  {Lynden-Bell}, {Yanny}, {Vidrih}, {Gilmore}, {Seabroke}, {Zucker},
  {Wilkinson}, {Hewett}, {Bramich}, {Fellhauer}, {Newberg}, {Wyse}, {Beers},
  {Bell}, {Barentine}, {Brinkmann}, {Cole}, {Pan}, \&
  {York}}]{Belokurov_et-al_2007}
---. 2007, \apj, 658, 337

\bibitem[{{Casey} {et~al.}(2012){Casey}, {Keller}, \& {Da
  Costa}}]{Casey_et-al_2012}
{Casey}, A.~R., {Keller}, S.~C., \& {Da Costa}, G. 2012, \aj, 143, 88

\bibitem[{{Fellhauer} {et~al.}(2007){Fellhauer}, {Evans}, {Belokurov},
  {Zucker}, {Yanny}, {Wilkinson}, {Gilmore}, {Irwin}, {Bramich}, {Vidrih},
  {Hewett}, \& {Beers}}]{Fellhaur_et-al_2007}
{Fellhauer}, M., {et~al.} 2007, \mnras, 375, 1171

\bibitem[{{Frebel} \& {Norris}(2011){Frebel}, \& {Norris}}]{Frebel_Norris_2011}
{Frebel}, A., \& Norris, J. E. 2011, arXiv/1102.1748

\bibitem[{{Grillmair} (2006)}]{Grillmair_2006}
{Grillmair}, C.~J. 2006, \apj, 643, L37

\bibitem[{{Hodapp} {et~al.}(2004){Hodapp}, {Kaiser}, {Aussel}, {Burgett},
  {Chambers}, {Chun}, {Dombeck}, {Douglas}, {Hafner}, {Heasley}, {Hoblitt},
  {Hude}, {Isani}, {Jedicke}, {Jewitt}, {Laux}, {Luppino}, {Lupton}, {Maberry},
  {Magnier}, {Mannery}, {Monet}, {Morgan}, {Onaka}, {Price}, {Ryan},
  {Siegmund}, {Szapudi}, {Tonry}, {Wainscoat}, \&
  {Waterson}}]{Hodapp_et-al_2004}
{Hodapp}, K.~W., {et~al.} 2004, Astronomische Nachrichten, 325, 636

\bibitem[{{Jin} \& {Lynden-Bell}(2007)}]{Jin_Lynden_Bell_2007}
{Jin}, S., \& {Lynden-Bell}, D. 2007, \mnras, 378, L64

\bibitem[{{Jordi} {et~al.}(2006){Jordi}, {Grebel}, \&
  {Ammon}}]{Jordi_et-al_2006}
{Jordi}, K., {Grebel}, E.~K., \& {Ammon}, K. 2006, \aap, 460, 339

\bibitem[{{Keller} {et~al.}(2007){Keller}, {Schmidt}, {Bessell}, {Conroy},
  {Francis}, {Granlund}, {Kowald}, {Oates}, {Martin-Jones}, {Preston},
  {Tisserand}, {Vaccarella}, \& {Waterson}}]{Keller_et-al_2007}
{Keller}, S.~C., {et~al.} 2007, PASA, 24, 1

\bibitem[{{Kerr} \& {Lynden-Bell}(1986)}]{Kerr_Lynden-Bell_1986}
{Kerr}, F.~J., \& {Lynden-Bell}, D. 1986, \mnras, 221, 1023

\bibitem[{{Koposov} {et~al.}(2010){Koposov}, {Rix}, \&
  {Hogg}}]{Koposov_et-al_2010}
{Koposov}, S.~E., {Rix}, H.-W., \& {Hogg}, D.~W. 2010, \apj, 712, 260

\bibitem[{{Marigo} {et~al.}(2008){Marigo}, {Girardi}, {Bressan}, {Groenewegen},
  {Silva}, \& {Granato}}]{Girardi_et-al_2008}
{Marigo}, P., {Girardi}, L., {Bressan}, A., {Groenewegen}, M.~A.~T., {Silva},
  L., \& {Granato}, G.~L. 2008, \aap, 482, 883

\bibitem[{{Mihalas} \& {Binney}(1981)}]{Mihalas_Binney_1981}
{Mihalas}, D., \& {Binney}, J. 1981, Science, 214, 829

\bibitem[{{Newberg} {et~al.}(2010){Newberg}, {Willett}, {Yanny}, \&
  {Xu}}]{Newberg_et-al_2010}
{Newberg}, H.~J., {Willett}, B.~A., {Yanny}, B., \& {Xu}, Y. 2010, \apj, 711,
  32

\bibitem[{{Odenkirchen} {et~al.}(2009){Odenkirchen}, {Grebel}, {Kayser}, {Rix},
  \& {Dehnen}}]{Odenkirchen_et-al_2009}
{Odenkirchen}, M., {Grebel}, E.~K., {Kayser}, A., {Rix}, H.-W., \& {Dehnen}, W.
  2009, \aj, 137, 3378

\bibitem[{{Robin} {et~al.}(2003){Robin}, {Reyl{\'e}}, {Derri{\`e}re}, \&
  {Picaud}}]{Robin_et-al_2003}
{Robin}, A.~C., {Reyl{\'e}}, C., {Derri{\`e}re}, S., \& {Picaud}, S. 2003,
  \aap, 409, 523

\bibitem[{{Roeser} {et~al.}(2010){Roeser}, {Demleitner}, \&
  {Schilbach}}]{Roeser_et-al_2010}
{Roeser}, S., {Demleitner}, M., \& {Schilbach}, E. 2010, \aj, 139, 2440

\bibitem[{{Sales} {et~al.}(2008){Sales}, {Helmi}, {Starkenburg}, {Morrison},
  {Engle}, {Harding}, {Mateo}, {Olszewski}, \& {Sivarani}}]{Sales_et-al_2008}
{Sales}, L.~V., {et~al.} 2008, \mnras, 389, 1391

\bibitem[{{Saviane} {et~al.}(2012){Saviane}, {Da Costa}, {Held}, {Sommariva},
  {Gullieuszik}, {Barbuy}, \& {Ortolani}}]{Saviane_et-al_2012}
{Saviane}, I., {Da Costa}, G.~S., {Held}, E.~V., {Sommariva}, V.,
  {Gullieuszik}, M., {Barbuy}, B., \& {Ortolani}, S. 2012, \aap, 540, A27

\bibitem[{{Schlegel} {et~al.}(1998){Schlegel}, {Finkbeiner}, \&
  {Davis}}]{Schlegel_Finkbeiner_Davis_1998}
{Schlegel}, D.~J., {Finkbeiner}, D.~P., \& {Davis}, M. 1998, \apj, 500, 525

\bibitem[{{Simon} {et~al.}(2011){Simon}, {Geha}, {Minor}, {Martinez},
	{Kirby}, {Bullock}, {Kaplinghat}, 
	{Strigari}, {Willman}, {Choi}, {Tollerud}, \&
	{Wolf}, J.}]{Simon_et-al_2011}
{Simon}, J.~D., {et~al.} 2011, \apj, 733, 46S

\bibitem[{{Starkenburg} {et~al.}(2010){Starkenburg}, {Hill}, {Tolstoy},
  {Gonz{\'a}lez Hern{\'a}ndez}, {Irwin}, {Helmi}, {Battaglia}, {Jablonka},
  {Tafelmeyer}, {Shetrone}, {Venn}, \& {de Boer}}]{Starkenburg_et-al_2010}
{Starkenburg}, E., {et~al.} 2010, \aap, 513, A34

\bibitem[{{Xue} {et~al.}(2008){Xue}, {Rix}, {Zhao}, {Re Fiorentin}, {Naab},
  {Steinmetz}, {van den Bosch}, {Beers}, {Lee}, {Bell}, {Rockosi}, {Yanny},
  {Newberg}, {Wilhelm}, {Kang}, {Smith}, \& {Schneider}}]{Xue_et-al_2008}
{Xue}, X.~X., {et~al.} 2008, \apj, 684, 1143

\bibitem[{{Zucker} {et~al.}(2006){Zucker}, {Belokurov}, {Evans}, {Kleyna},
  {Irwin}, {Wilkinson}, {Fellhauer}, {Bramich}, {Gilmore}, {Newberg}, {Yanny},
  {Smith}, {Hewett}, {Bell}, {Rix}, {Gnedin}, {Vidrih}, {Wyse}, {Willman},
  {Grebel}, {Schneider}, {Beers}, {Kniazev}, {Barentine}, {Brewington},
  {Brinkmann}, {Harvanek}, {Kleinman}, {Krzesinski}, {Long}, {Nitta}, \&
  {Snedden}}]{Zucker_et-al_2006}
{Zucker}, D.~B., {et~al.} 2006, \apjl, 650, L41





\end{thebibliography}
\end{document}